\documentclass[%
 reprint,
 showpacs,
 showkeys,
 amsmath,
 amssymb,
 aps,
 prl,
 floatfix,
]{revtex4-1}

\usepackage{graphicx}					
\usepackage[makeroom]{cancel}				

\newcommand{\pd}{\partial}				
\newcommand{\dd}{\mathrm{d}}				
\newcommand{\K}{\mathcal{K}}
\newcommand{\h}{\mathrm{H}}

\newcommand{\M}{\mathrm{M}}
\newcommand{\ads}{\mathrm{arcds}}

\begin{document}

\title{Rotation number of integrable symplectic mappings of the plane}
\author{Timofey Zolkin}
\email{zolkin@fnal.gov}
\affiliation{Fermilab, Batavia, IL 60510, USA}
\author{Sergei Nagaitsev}
\affiliation{Fermilab, Batavia, IL 60510, USA}
\affiliation{Department of Physics, The University of Chicago, Chicago, IL 
60637, USA}
\author{Viatcheslav Danilov}
\email{Deceased}
\affiliation{Oak Ridge National Laboratory, Oak Ridge, TN 37831, USA}
\date{\today}

\begin{abstract}

Symplectic mappings are discrete-time analogs of Hamiltonian systems.
They appear in many areas of physics, including, for example, accelerators, 
plasma, and fluids.  
Integrable mappings, a subclass of symplectic mappings, are equivalent to a 
Twist map, with a rotation number, constant along the phase trajectory.
In this letter, we propose a succinct expression to determine the rotation 
number and present two examples.
Similar to the period of the bounded motion in Hamiltonian systems, the 
rotation number is the most fundamental property of integrable maps and it 
provides a way to analyze the phase-space dynamics.

\end{abstract}
                             

\maketitle


For a one degree-of-freedom time-independent system, the Hamiltonian 
function, $\h[p,q;t] = E$, is the integral of the motion.
If the motion is bounded, it is also periodic and the period of oscillations can
be determined by integrating
\begin{equation}
\label{math:T}
	T = \oint\left(\frac{\pd \h}{\pd p}\right)^{-1}\dd q,
\end{equation}
where $p=p(E, q)$. 
Similarly, a map $(q',p') = \M(q,p)$ in the plane is called {\it integrable}, 
if 
there is a
non-constant real-valued continuous function $\K(q,p)$, which is invariant 
under $\M$.
The function $\K(q,p)$ is called {\it integral}.
In this paper, we are describing the case, for which the level sets
$\K = \text{const}$ are compact closed curves (or sets of points) and for which 
the identity
\[
	\K(q',p') = \K(q,p)
\]
holds for all $(q,p)$.
There are many examples of integrable mappings, including the famous McMillan
mapping \cite{mcmillan1971problem}, described below. The dynamics is in many 
ways similar to that of a continuous system, however, Eq.~(\ref{math:T}) is 
not 
directly applicable since the integral $\K(q,p)$ is not the Hamiltonian 
function.

The {\bf Arnold-Liouville theorem} for maps 
\cite{veselov1991integrable,arnold1968ergodic} states that in action-angle 
variables, consecutive iterations of map $\M$ lie on nested circles of radius 
$J$ and that the map can be written in the form of a Twist map
\begin{eqnarray}
J_{n+1}		&=& J_n,				\\
\theta_{n+1}	&=& \theta_n + 2\,\pi\,\nu(J) \mod 2\,\pi,
\end{eqnarray}
where $|\nu(J)|\leq0.5$ is the rotation number, $\theta$ is the angle variable 
and $J$ is the action variable, defined by the mapping $\M$ as
\begin{equation}
\label{math:J}
	J = \frac{1}{2\,\pi} \oint p\,\dd q.
\end{equation}

For integrable mappings, $\K(q,p) = \K(J)$ is a function of the action variable.
In what follows, we present a simple analytical expression to calculate the 
rotation number, $\nu(\K)$, without constructing an action-angle transformation.
This is useful, when, for example, the action variable~(\ref{math:J})
is not known explicitly but an integral $\K(q,p)$ is.

{\bf Theorem (Danilov):}
\begin{equation}
\label{math:Danilov}
\nu(\K) =	\int_{q}^{q'}
	\left(\frac{\pd\K}{\pd p}\right)^{-1}
	\dd q
	\Bigg/
	\oint
		\left(\frac{\pd\K}{\pd p}\right)^{-1}
	\dd q,						\\
\end{equation}
where both integrals are taken along the invariant curve, $\K(q, p)$.

\begin{figure}[th!]\centering
\includegraphics[width=\linewidth]{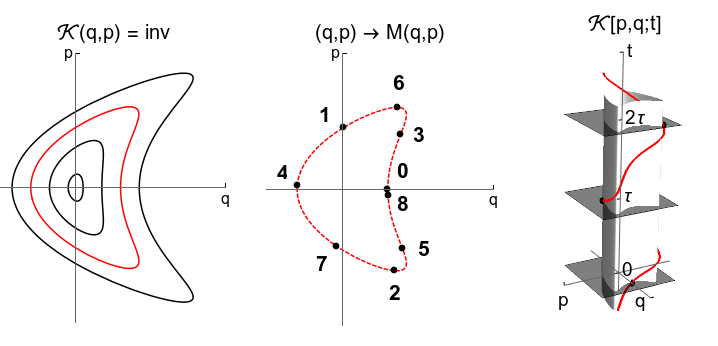}
\caption{\label{fig:DanilovTh}
	Constant level sets of the integral $\K(q,p)=\text{const}$ (left).
	A particular curve representing a level set of $\K$
	and several iterates of the map $\M$ (center).
	A three-dimensional phase space, $(q,p)$ + time, of the 
	system~(\ref{math:Kham}) (right).
	Dark gray planes $t=0,\tau,2\tau,\ldots$ represent stroboscopic 
	Poincar\'e section of the continuous flow of the system 
	(red curve) which is identical to map $\M$.
	}
\end{figure}

\noindent
{\bf Proof:} 
Consider the following system of differential equations:
\begin{equation}
\label{math:Kham}
	\frac{\dd\,q}{\dd t} =  \frac{\pd\K}{\pd p}, \qquad
	\frac{\dd\,p}{\dd t} = -\frac{\pd\K}{\pd q},
\end{equation}
such that $\K(q,p)$ does not change along a solution of the system.  
Define a new map, $\widetilde{\M}(q,p)$ (see Fig.~\ref{fig:DanilovTh})
\begin{equation}
(q',p') = \widetilde{\M} (q,p) = \left(q(\tau),p(\tau)\right)
\end{equation}
with
\begin{equation}
	q = q(0)\quad\text{and}\quad p = p(0),
\end{equation}
where $q(t)$ and $p(t)$ are the solutions of the system~(\ref{math:Kham}) and 
$\tau$ is the discrete time step. 
For a given value of $\K$, which is an integral of both $\M$ and 
$\widetilde{\M}$, one can always select $\tau(\K)$ such that the maps 
$\M(q,p)$ and $\widetilde{\M}(q,p)$ are identical.
Since $\K(q,p)$ is compact and closed, the functions $q(t)$ and $p(t)$ are 
periodic with a period $T(\K)$.
By its definition,
\begin{equation}
\tau = \nu(\K)T(\K).
\end{equation}
Let us now calculate $\nu(\K)$:
\begin{equation}
\nu(\K) \equiv
\frac{\tau}{T} = 
\frac{\int_{q}^{q'}	\,\dd t}
     {\oint		\,\dd t} =
\frac{\int_{q}^{q'}	\left( \frac{\dd\,q}{\dd t} \right)^{-1}\,\dd q}
     {\oint		\left( \frac{\dd\,q}{\dd t} \right)^{-1}\,\dd q} =
\frac{\int_{q}^{q'} \left(\frac{\pd \K}{\pd p}\right)^{-1}\,\dd q}
     {\oint         \left(\frac{\pd \K}{\pd p}\right)^{-1}\,\dd q}.
\end{equation}
Q.E.D..

In order to employ this theorem in practice, one would need to recall that with
$p=p(\K,q)$, the integrand in Eq.~(\ref{math:Danilov}),
\[
	\left( \frac{\pd\K}{\pd p} \right)^{-1},
\]
is the function of only $q$ for a given $\K = \text{const}$.
Also, the lower limit of the integral can be chosen to be any convenient value 
of $q$, for example 0, as long it belongs to a given level set, $\K(q,p)$.
Finally, the upper limit of the integral, $q'$, is obtained from the selected
$q$ and $p=p(\K,q)$ by the map, $\M(q,p)$.
Let us now consider several examples.


As our first example, we will consider a linear symplectic map,
\begin{equation}
\label{math:LMap}
\begin{bmatrix}
q'	\\ p'
\end{bmatrix}
=
\begin{bmatrix}
 a & b		\\
 c & d
\end{bmatrix}
\begin{bmatrix}
q	\\ p
\end{bmatrix},
\end{equation}
with $a\,d-b\,c=1$ and $|a+d| \leq 2$.
This mapping is very common in accelerator physics and has been described in 
\cite{courant1958theory}.
The rotation number for this mapping is well known:
\begin{equation}
	\nu = \frac{1}{2\,\pi}\,\arccos \frac{a+d}{2}.
\end{equation}
In order to employ the Danilov theorem, we will use the following 
parametrization:
\begin{eqnarray}
a-d &=& 2\,\alpha\,\sin(2\,\pi\nu),		\\
b &=& \beta\,\sin(2\,\pi\nu),			\\
c &=& -\gamma\,\sin(2\,\pi\nu).
\end{eqnarray}
The symplecticity condition gives $\beta\,\gamma-\alpha^2=1$.
With this parametrization, an integral of mapping~(\ref{math:LMap}) can be 
written as
\begin{equation}
	\K = \gamma\,q^2 + 2\,\alpha\,q\,p + \beta\,p^2.
\end{equation}
To calculate the rotation number, we first express $p$ through $\K$ and $q$:
\begin{equation}
p = \frac{-\alpha\,q \pm \sqrt{\alpha^2q^2-\beta(\gamma\,q^2-\K)}}{\beta}
  = \frac{-\alpha\,q \pm \sqrt{\beta\,\K-q^2}}{\beta}.
\end{equation}
Now
\begin{equation}
\left( \frac{\pd\K}{\pd p} \right)^{-1} =
\frac{1}{2(\alpha\,q+\beta\,p)} = 
\frac{\pm 1}{2\sqrt{\beta\,\K-q^2}}.
\end{equation}
We will use
\begin{equation}
(q,p) = (0,\sqrt{\K/\beta})
\end{equation}
and
\begin{equation}
(q',p') = (b\,\sqrt{\K/\beta},d\,\sqrt{\K/\beta})
\end{equation}
to evaluate the integral in the numerator:
\begin{eqnarray}
\int_{0}^{b\,\sqrt{\K/\beta}}
	\frac{\dd q}{2\sqrt{\beta\,\K-q^2}} & = &
\int_{0}^{b/\beta}
	\frac{\dd x}{2\sqrt{1-x^2}}		\nonumber\\
& = &
\frac{1}{2} \arcsin \frac{b}{\beta} =
\pi\,\nu.
\end{eqnarray}
The integral in the denominator equals $\pi$.
Thus, the rotation number is $\nu$.


As our second example, we will consider the so-called McMillan map 
\cite{mcmillan1971problem},
\begin{equation}
\label{math:McMap}
\begin{bmatrix}
q'	\\ p'
\end{bmatrix}
=
\begin{bmatrix}
p	\\ -q + \frac{2\,\epsilon\,p}{p^2+\Gamma}
\end{bmatrix}.
\end{equation}
To illustrate the Danilov theorem, we will limit ourselves to a case with
$\Gamma > 0$ and $|\epsilon| \leq \Gamma$.
Mapping~(\ref{math:McMap}) has the following integral:
\[
	\K(q,p) = q^2 p^2 + \Gamma(q^2+p^2) - 2\,\epsilon\,q\,p,
\]
which is non-negative for the chosen parameters.

We first notice that for small amplitudes $p^2 \ll \Gamma$, the rotation number 
is
\begin{equation}
	\nu \approx \frac{1}{2\,\pi}\,\arccos\frac{\epsilon}{\Gamma},
\end{equation}
while at large amplitudes, the rotation number becomes $0.25$.
Again, we first express $p$ through $\K$ and $q$ and evaluate the integrand
in~(\ref{math:Danilov}):
\begin{equation}
\left( \frac{\pd \K}{\pd p} \right)^{-1} =
\frac{1}{2\,p\,(q^2+\Gamma)-2\,\epsilon\,q} =
\frac{\pm 1}{2\sqrt{-\Gamma\,q^4+\delta\,q^2+\lambda\,\K}},
\end{equation}
where $\delta(\K) = \epsilon - \Gamma^2 +\K$.
Let us define a parameter,
\begin{equation}
k(\K) = 
\frac{1}{\sqrt{2}}\sqrt{1+\frac{\delta}{\sqrt{\delta^2+4\,\K\,\Gamma^2}}},
\end{equation}
which spans from 0 to 1.
Also, define $k' = \sqrt{1-k^2}$.
Then, the rotation number can be expressed through Jacobi elliptic functions as 
follows:
\begin{equation}
\nu(\K) = \frac{1}{4\,\mathrm{K}(k)}
	\ads \left( \sqrt{\frac{k\,k'\,\Gamma}{\sqrt{\K}}} , k \right),
\end{equation}
where $\mathrm{K}(k)$ is the complete elliptic integral of the first kind and 
the inverse Jacobi function, $\ads(x,k)$, is defined as follows
\begin{equation}
\ads(x,k) = \int_x^\infty \frac{\dd t}{\sqrt{(t^2+k^2)(t^2-k'^2)}}.
\end{equation}

\begin{figure}[t!]\centering
\includegraphics[width=\linewidth]{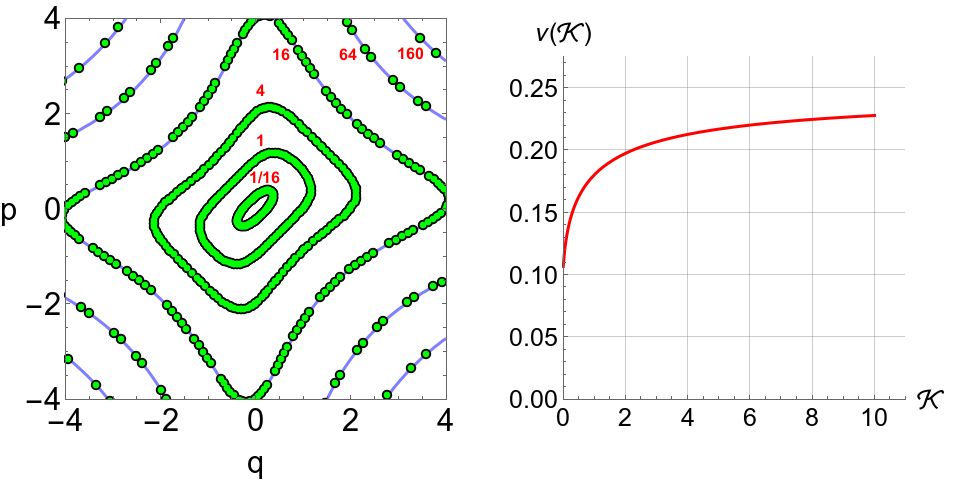}
\caption{\label{fig:McM}
	Left plot contain various graphical iterations of the canonical 
McMillan 
	map ($\epsilon = 0.8$, $\Gamma=1$).
	Constant level sets of the invariant are shown with blue lines and
	corresponding value of the invariant $\K$ is shown in red.
	Right plot is the rotation number as a function of its integral.
	} 
\end{figure}

Figure~\ref{fig:McM} shows an example of the rotation number, for the case of 
$\epsilon=0.8$
and $\Gamma=1$ ($\nu(0)\approx 0.102$), as a function of integral, $\K$.

These two examples demonstrate that the Danilov theorem is a powerful tool.
The McMillan map is a classic example of a nonlinear integrable discrete-time 
system.
It is a typical member of a wide class of area-preserving transformations 
called a Twist map \cite{meiss1992symplectic}.
In this Letter we demonstrated a general and exact method on how to find a 
Poincar\'e rotation number.
It complements the discrete Arnold-Liouville theorem for maps 
\cite{veselov1991integrable,arnold1968ergodic} 
and permits the analysis of the system dynamics.
In conclusion, we would like to point out that for cases when the integral $\K$
is also known as a function of action, $J$, one would be able to express the 
rotation number as a function of action, as well as the Hamilton’s function, 
$\mathrm{H}(J)$, for mapping since
\begin{equation}
	\nu(J) = \frac{\dd\,\mathrm{H}}{\dd J}.
\end{equation}

This work has been partially supported by the NSF Grants PHY-1535639 
and PHY-1549132.  Fermilab is Operated by Fermi Research Alliance, LLC under 
Contract
No.~De-AC02-07CH11359 with the U.S. Department of Energy.


\providecommand{\noopsort}[1]{}\providecommand{\singleletter}[1]{#1}%


\end{document}